# Digital Business Model Analysis Using a Large Language Model

Masahiro Watanabe[*], Naoshi Uchihira[*]

**Abstract**

Digital transformation (DX) has recently become a pressing issue for many companies as the latest digital technologies, such as artificial intelligence and the Internet of Things, can be easily utilized. However, devising new business models is not easy for companies, though they can improve their operations through digital technologies. Thus, business model design support methods are needed by people who lack digital technology expertise. In contrast, large language models (LLMs) represented by ChatGPT and natural language processing utilizing LLMs have been developed revolutionarily. A business model design support system that utilizes these technologies has great potential. However, research on this area is scant. Accordingly, this study proposes an LLM-based method for comparing and analyzing similar companies from different business domains as a first step toward business model design support utilizing LLMs. This method can support idea generation in digital business model design.

## 1 Introduction

With the advancement of digital technology, companies are compelled to reassess their business operations fundamentally. Digital transformation (DX) is crucial for maintaining sustainable firm growth and competitiveness. Concurrently, natural language processing (NLP) technologies based on large language models (LLMs) have significantly advanced. LLM-based applications are expected to be utilized increasingly. They include ChatGPT, which understands and generates human-like text. However, the use of LLMs for analyzing and designing new business models is still evolving. Particularly, research related to LLM-based analysis and support for digital business model design is scarce.

Thus, this study aims to use an LLM to analyze documents detailing various companies' DX initiatives, focusing on the similarities of their digital business models to aid in designing new models. In the reports [1] of Japan's Ministry of Economy, Trade and Industry (METI), known as "DX Stocks" (formerly "Proactive IT Management Stocks"), METI has been selecting DX-promoting firms to disseminate exemplary digital business models since 2015. We used these documents for analysis, which include over 350 DX examples documented. By applying an LLM to measure the similarity of documents, DX designers can consider new business models by referring to similar business models.

For the rest of this paper, Section 2 briefly reviews related literature on business model analysis and design. Section 3 describes the methodology used in our study. Section 4 presents the analysis examples, and Section 5 presents the discussion and conclusion.

---

[*] Japan Advanced Institute of Science and Technology, Ishikawa, Japan

## 2  Literature Review

Many analytical approaches to business models are available [2]. Although less common than analytical approaches, research on business model design is also gaining momentum. Business model design methods include the Business Model Canvas (BMC) [3] and the Business Model Navigator [4], for which many proposals for extension are suggested.

BMC is a generic framework, though it can be patterned for each business domain of interest. Studies on BMC patterns for the Internet of Things (IoT) include Dijkman et al. [5] and Ju et al. [6]. Uchihira et al. [7] proposed the BMC design pattern for IoT businesses. In these studies, researchers extracted patterns manually. The process requires considerable effort and time by researchers and cannot keep up with the latest situation where new business models using digital technology are appearing consecutively. In contrast, numerous business analyses use text mining. In recent years, NLP using LLMs has evolved dramatically and is outperforming the limits of conventional text mining. The manual analysis by researchers is expected to be automated by NLP using LLMs.

Corporate analysis using NLP is predominantly conducted with a focus on economic trends and stock price prediction, with studies such as those by Taniguchi et al. [8] and Doi et al. [9]. Additionally, Gurcan et al. used Latent Dirichlet Allocation (LDA) to identify 34 topics from papers on DX published over the past decade, examining the strategies, practices, and trends of DX [10]. However, studies analyzing DX-related texts to support digital business model designs are extremely rare.

## 3  Analysis Method

This study analyzes the report texts of DX stocks designated by METI since 2015. These reports highlight listed companies and aim to promote exemplary business models that serve as benchmarks for DX, showcasing advanced digital business practices.

The specific analysis methods used in this study include the following (Figure 1):

1. Preprocessing of report texts (removal of stop words, normalization).
2. Vectorizing the texts using "LUKE" [11], a cutting-edge Japanese pretrained LLM, to process the texts numerically.
3. Selecting one DX case of the reference company A.
4. Calculating cosine similarity to measure the similarity between the DX case of company A and those of different companies in different business domains.
5. Selecting two companies B and C with the highest similarity scores to the case of company A and analyzing the common points and useful aspects of DX for reference.

Cosine similarity is used to measure the similarity between two vectors as represented by the following formula:

$$Cosin\ Similarity(A, B) = \frac{A \cdot B}{\|A\|\|B\|}$$

"LUKE" is a language model developed on the transformers architecture, specifically designed to excel in language understanding tasks with enhanced entity recognition capabilities. This model is particularly adept at identifying proper nouns and technical terms within texts and analyzing their relationships in detail. It is especially effective for examining textual data pertinent to companies' digital strategies and technological implementations.

This study aims to support the generation of ideas for digital business models in the future and marks the first step in that direction. By learning patterns from the commonalities of DX cases, companies can use this knowledge as a reference when considering their DX initiatives.

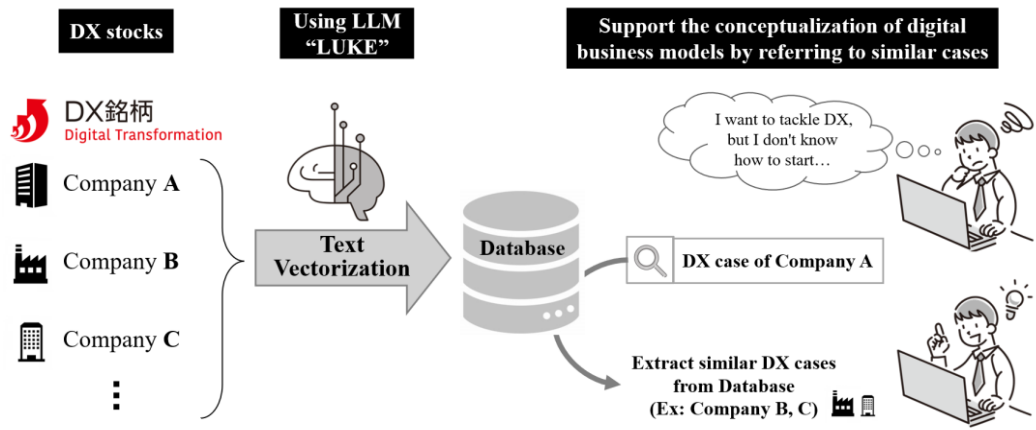

Figure 1: Proposed Analysis Method

## 4 Analysis Examples

Using the "LUKE" language model, the similarity of reports from various companies can be assessed quantitatively, revealing instances where companies across different industries have adopted similar DX strategies. To support business model designs, the effectiveness of company similarity is evaluated using the following steps:

1. Select one target company A.
2. Retract two reference companies B and C from different business domains, which are similar to A based on cosine similarity among vectorized tests.
3. Interpret the similarity between A, B, and C from a perspective that can be used as a reference when considering new business models.

### 4.1 Analysis Example 1 (Table 1):

A) Target Company A: pharmaceutical company
B) Reference Company B: beverage manufacturer
C) Reference Company C: chemical manufacturer

Table 1: Similarity of cases to pharmaceutical company A

| Company | Industry classification | Cos Similarity | Common Features |
|---|---|---|---|
| B | Manufacturing | 0.954170 | AI, communication platform, |
| C | Manufacturing | 0.951046 | modernized, Making efficient |

**Pharmaceutical Company A**
1. AI for predictive accuracy in management
2. A new platform for sharing ideas from the field
3. Modernized 50-year-old system
4. Overhauled global core IT systems

**Beverage Manufacturer B**
1. AI and robotic process automation (RPA) for problem-solving and value creation
2. Optimized IT for global competitiveness
3. Communication promotion for collaboration between business and IT departments
4. Updated beer quality system for improved efficiency and consistent quality

**Chemical Manufacturer C**
1. AI usage to enhance sales processes and develop new ventures
2. Collaboration with overseas affiliates and promotion of diverse team interactions
3. Core systems updated to the public cloud and maintenance outsourced
4. Improved data management and security through digitization and tablet use in the field

**Similarity Analysis**

These globally operating companies use data-driven management to stay competitive in international markets and enter new ones. They actively develop communication platforms to share insights and strategies at the operational level, including personnel exchanges. This strategy shows a strong commitment by global manufacturers to build digital strategies and foster a transformative mindset throughout the organization.

**Points of Reference for DX**
- AI usage: for predictive accuracy, problem-solving, and enhancing sales
- New communication platforms: for idea-sharing and collaboration
- System modernization: to update old systems and overhaul core systems

### 4.2 Analysis Example 2 (Table 2):

A) Target Company A: chemical manufacturer
B) Reference Company B: air conditioning manufacturer
C) Reference Company C: trading company

Table 2: Similarity of cases against chemical manufacturer A

| Company | Industry classification | Cos similarity | Common features |
|---|---|---|---|
| B | Manufacturing | 0.953263 | DX center, COVID-19, AI, Making efficient |
| C | Wholesale | 0.948659 | |

**Chemical Manufacturer Company A**
1. Advanced technology strategy office for digital initiatives
2. AI for new value creation
3. Reduced manual labor and streamlined processes
4. Work-at-home system for the COVID-19 pandemic

**Air Conditioning Manufacturer B**
1. Technology innovation center for digital adoption
2. RPA and chatbots for efficiency and cost-cutting
3. Global initiative: engaging in global digital projects such as smart cities
4. COVID-19 task force for sales strength and resilience

**Trading Company C**
1. DX center for new business and innovation
2. Global DX initiatives with smart city projects in Vietnam
3. RPA for new services and revenue
4. Flexible work styles for the COVID-19 pandemic

**Similarity Analysis**

A key trait among these companies is the formation of specialized digital departments within their DX strategies. Establishing dedicated DX units is a crucial step toward facilitating smooth organizational transformation. Moreover, they are utilizing digital technologies to construct global digital business models and innovate using AI. Additionally, their adaptive use of digital tools in response to the COVID-19 pandemic highlights a common strategic approach across these firms.

**Points of Reference for DX**
- Specialized DX departments: establishing dedicated digital units
- Global initiative: engaging in global digital projects such as smart cities
- Process automation: implementing RPA and chatbots
- COVID-19 adaptations: flexible work styles and task forces

## 5  Discussion and Conclusion

This study utilized an LLM to calculate the cosine similarity between various case studies described in reports on DX. By quantitatively assessing the similarity to specific cases, we uncovered detailed insights into DX strategies that encompass specific business processes and technological innovations. Two analysis examples show that LLM can effectively extract similar DX cases. Specifically, by referencing similar cases not only within the same industry but also from different industries and considering their commonalities, the ideation of digital business models can be supported. Moreover, this result would not be possible with a simple keyword search, and we confirmed the effectiveness of LLMs in supporting business model design. LLMs can find similarities in the meaning of sentences compared with traditional keyword searches and text mining. The proposed method also potentially offers companies easy access to insights into the use of digital technologies and business model innovations that have previously been less accessible.

Thus far, little research applies NLP and LLMs to business model analysis to support business model ideation. Although the results of this study are preliminary, they make certain academic contributions by demonstrating the potential of this approach.

In the future, we plan to refine our analytical methods using advanced NLP technologies and broaden our examination of digital business models across a wider spectrum of industries. Furthermore, we are considering the development of a recommendation system, possibly implemented via chatbots, that could suggest similar cases. This system is expected to act as a catalyst for companies aiming to accelerate their DX efforts, providing them with essential guidance as they tackle the challenges of DX. We have also proposed the digital innovation design method [12]. The recommendation system using an LLM can be incorporated into this design method.